\documentclass[%
prl,
 reprint,
 amsmath,amssymb,
 aps,preprintnumbers]{revtex4-1}

\usepackage[colorlinks=true,linkcolor=black, citecolor=black,
urlcolor=black]{hyperref}

\usepackage{multirow,graphics}
    \newcommand{\beq}{\begin{equation}}
    \newcommand{\eeq}{\end{equation}}
    \newcommand\beqa{\begin{eqnarray}}
    \newcommand\eeqa{\end{eqnarray}}

\usepackage{amstext}
\usepackage{amssymb}
\usepackage{amsmath}
\usepackage{graphicx}
\usepackage{color}

\begin{document}

\newcommand{\IM}{{\rm Im}\,}
\newcommand{\card}{\#}
\newcommand{\la}[1]{\label{#1}}
\newcommand{\eq}[1]{(\ref{#1})}
\newcommand{\figref}[1]{Fig \ref{#1}}
\newcommand{\abs}[1]{\left|#1\right|}
\newcommand{\comD}[1]{{\color{red}#1\color{black}}}

\makeatletter
     \@ifundefined{usebibtex}{\newcommand{\ifbibtexelse}[2]{#2}} {\newcommand{\ifbibtexelse}[2]{#1}}
\makeatother

\preprint{Imperial/TP/13/SL/02}

\newcommand{\footnoteab}[2]{\ifbibtexelse{%
\footnotetext{#1}%
\footnotetext{#2}%
\cite{Note1,Note2}%
}{%
\newcommand{\textfootnotea}{#1}%
\newcommand{\textfootnoteab}{#2}%
\cite{thefootnotea,thefootnoteab}}}
\newcommand{\footnoteb}[1]{\ifbibtexelse{\footnote{#1}}{%
\newcommand{\textfootnoteb}{#1}%
\cite{thefootnoteb}}}
\newcommand{\footnotebis}{\ifbibtexelse{\footnotemark[\value{footnote}]}{%
\cite{thefootnoteb}}}

\def\e{\epsilon}
     \def\bT{{\bf T}}
    \def\bQ{{\bf Q}}
    \def\wT{{\mathbb{T}}}
    \def\wQ{{\mathbb{Q}}}
    \def\ttQ{{\bar Q}}
    \def\tQ{{\tilde \bP}}
        \def\bP{{\bf P}}
    \def\CF{{\cal F}}
    \def\cC{\CF}
     \def\Tr{\text{Tr}}
     \def\l{\lambda}
\def\hbZ{{\widehat{ Z}}}
\def\bZ{{\resizebox{0.28cm}{0.33cm}{$\hspace{0.03cm}\check {\hspace{-0.03cm}\resizebox{0.14cm}{0.18cm}{$Z$}}$}}}
\newcommand{\rb}{\right)}
\newcommand{\lb}{\left(}

\newcommand{\gT}{T}\newcommand{\gQ}{Q}

\title{Quantum Spectral Curve  for \texorpdfstring{AdS\(_5\)/CFT\(_4\)}{AdS\_5/CFT\_4}}

\author{ Nikolay Gromov$^{a}$, Vladimir Kazakov$^{b}$, S\'ebastien Leurent$^{c}$, Dmytro Volin$^{d}$}

\affiliation{%
\(^{a}\)Mathematics Department, King's College London, The Strand, London WC2R 2LS, UK \&
 St.Petersburg INP, Gatchina, 188300, St.Petersburg, Russia
\\
             \(^{b}\) LPT, \'Ecole Normale Superieure, 24, rue Lhomond 75005 Paris, France \& Universit\'e Paris-VI, Place Jussieu, 75005 Paris, France\\
 \(^{c}\) Imperial College,
London SW7 2AZ, United Kingdom
\\
               \(^{d}\) Nordita, KTH Royal Institute of Technology and Stockholm University, Roslagstullsbacken 23, SE-106 91 Stockholm, Sweden
               }

\begin{abstract}
We present a new formalism, alternative to
the old TBA-like approach, for  solution of the spectral
problem of planar \({\cal N}=4\) SYM.
 It takes
a concise form of a non-linear  matrix Riemann-Hilbert problem in
terms of a few Q-functions.
We demonstrate the formalism for two types of observables -- local operators
at weak coupling and cusped Wilson lines in a near BPS limit.
\end{abstract}

 \maketitle

\section{Introduction}
\begin{figure}
  \centering
  \includegraphics{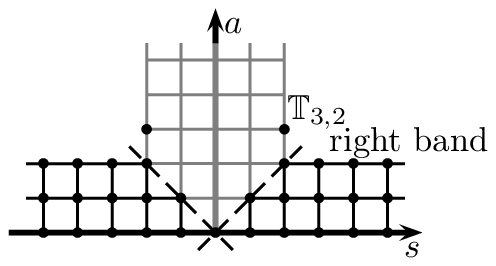}
  \caption{T-hook: lattice for the AdS/CFT T-system}
  \label{fig:THook}
\end{figure}

The  spectrum of anomalous dimensions in the planar
\({\cal N}=4\) SYM theory  was successfully  studied in the last decade, to great extent due to the ideas of AdS/CFT correspondence  and integrability \cite{Beisert:2010jr}.  A conventional form of  solution to the spectral problem
 is given by an infinite set
of nonlinear integral TBA equations \cite{Bombardelli:2009ns,Gromov:2009bc,Arutyunov:2009ur} for the functions of the spectral parameter \(Y_{a,s}(u)\)
\begin{equation*}
\log Y_{as}(u)=\delta_{s}^0\, iLp_a(u)+\int dv K_{as}^{a'\!s'}\!(u,v)\!\log(1+Y_{a's'}(v))
\end{equation*}
where the sum over \(a',s'\) in the
 r.h.s. goes along the  {internal} nodes of the lattice (T-hook) in \figref{fig:THook}. The momentum \(p_a\)  and the kernels
 \(K_{as}^{a's'}\) are explicit but rather complicated functions of  the spectral parameters  \(u,v\) \cite{Gromov:2009bc}. Their important analytic feature is the presence of cuts, parallel to \(\mathbb{R}\),  with  fixed branch-points at \(u,v\in\pm2 g+i\mathbb{Z}\) or \(u,v\in\pm
 2 g+i\mathbb{(Z} +\frac{1}{2})\) where  \(g\equiv \sqrt \l /(4\pi)\)  and  \(\l\) is the 't~Hooft coupling.
This TBA system fixes completely the Y-functions and hence the dimension of a particular operator
specified by certain poles and zeros incorporated into the driving terms  \cite{Gromov:2009bc}.
It was successfully used for the weak and strong coupling analysis \cite{Arutyunov:2010gb,Balog:2010vf,Balog:2010xa} as well as  for the first successful numerical computations of dimensions of Konishi \cite{Gromov:2009zb,Frolov:2010wt} and similar  operators \cite{Gromov:2011de,Frolov:2012zv}.
However, this TBA system has very complex analyticity properties, which  limits in practice its applications and obscures the long anticipated beauty of the whole problem.

An obvious sign of this  hidden beauty is the direct equivalence  of the TBA system to the   AdS/CFT Y-system,
originally proposed as a solution of the AdS/CFT spectral problem in \cite{Gromov:2009tv}, with  additional analyticity conditions  \cite{Cavaglia:2010nm}. It is a universal set of equations equivalent, by the
substitution \(Y_{a,s}=\frac{{\mathbb T}_{s+1,a}{\mathbb
    T}_{s-1,a}}{{\mathbb T}_{a+1,s}{\mathbb T}_{a-1,s}}\), to the
Hirota discrete bilinear equation (T-system) \footnoteab{We denote  \(f^\pm=f(u\pm \tfrac i2)\) and
    \(f^{[\pm k]}=f(u\pm k\tfrac i2)\).}{With the present definition
    of  ${\mathbb T}_{a,s}$, \eqref{T-system}
  holds when $|\mathrm{Im}(u)|<s-a$, and an analytic continuation is
  necessary to make \eqref{T-system} hold everywhere by imposing long cuts
  for ${\mathbb T}_{a,s}$.}
\begin{equation}
\wT_{a,s}^+\wT_{a,s}^-=\wT_{a+1,s}\wT_{a-1,s}
+\wT_{a,s+1}\wT_{a,s-1},
\label{T-system}\end{equation}
which is integrable in its
turn.Using this integrability the general solution of  T-system
 can be  explicitly parameterized in terms of Wronskians
 built from only 8 independent Q-functions \cite{Gromov:2010vb,Gromov:2010km}.
 The Q-functions
are the most elementary constituents of the whole
   construction with the  analyticity properties much simpler than those of
   Y- or T-functions  \cite{Gromov:2011cx}. With a savvy choice of the basic Q-functions we managed in
   \cite{Gromov:2011cx} to close a finite system of non linear
   integral equations (FiNLIE).
   It appeared to be an very efficient tool  in multi-loop weak coupling computations \cite{Leurent:2012ab,Leurent:2013mr}. But it was clear that the somewhat bulky form of that FiNLIE  \cite{Gromov:2011cx} hides a much more beautiful and simple formulation, with a clear insight into the full analytic structure of the underlying functions.

 We formulate in this note a new, much more transparent and  concise system of the planar AdS\(_5\)/CFT\(_4\)     spectral equations   of the Riemann-Hilbert type.  It might represent the ultimate simplification of this spectral problem.

\section{\texorpdfstring{\(\bP\mu\)}{P-mu} system for the spectrum}

We will demonstrate our new approach on the most important example of the left-right symmetric states for which \({\wT}_{a,s}={\wT}_{a,-s}\) (in the appropriate  gauge described in \cite{Gromov:2011cx}).
To start with, all T- and Y-functions
can be expressed in terms of  \(4+4\) Q-functions \cite{Gromov:2010km}.
Let us exemplify this relation for  T-functions of the right band (see
\figref{fig:THook}),
where we have for \(s>0\) \cite{Gromov:2011cx}
\begin{equation}\label{Tsr}
{\mathbb T}_{1,s}(u)=\bP_1(u+\tfrac{is}{2})\bP_2(u-\tfrac{is}{2})
- \bP_2(u+\tfrac{is}{2})\bP_1(u-\tfrac{is}{2}),
\end{equation}
where the symbol \(\bP\) is used to denote the Q-functions in the right
band, in order  to avoid a clash with other notations existing in the literature.
An important feature of this parameterization is that \(\bP\)'s
have only one single cut between  \(-2g\) and \(2g\),  otherwise  being analytic
in the whole complex plane \footnoteb{More precisely, it is in general the square \(\bP_i^2\) of these
  Q-functions which has a single cut.
For operators with half-integer asymptotic behaviour,
\(\bP_a\) has an additional quadratic branch point at infinity.
This branch point is absent from the physical quantities, as they are expressed through products and
  ratios of two \(\bP\)-s (cf. \eqref{Y11Y22}).} \cite{Gromov:2011cx}. This property is tightly related to what we refer to as  \({\mathbb Z}_4\)-symmetry \cite{Bena:2003wd,Beisert:2005bm,Gromov:2011cx}.
\begin{figure}
  \centering
  \includegraphics[scale=0.6]{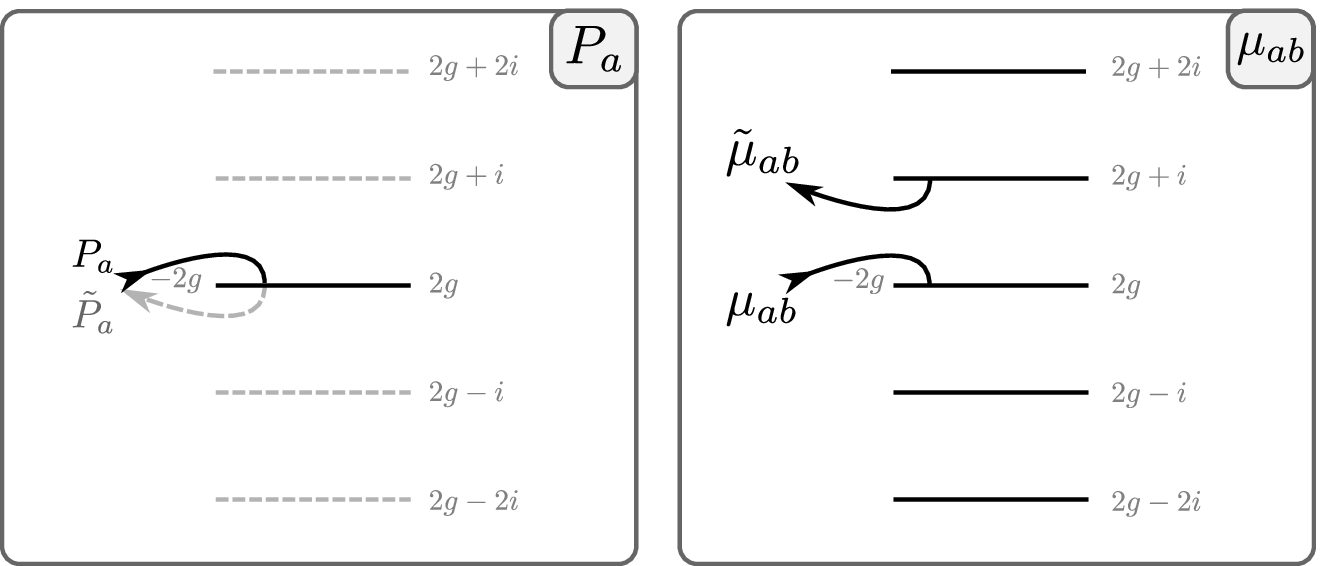}
  \caption{Cut structure of \(\bP\) and \(\mu\)}
  \label{fig:Cuts}
\end{figure}

Ideally, we would like to reduce the whole problem to a single spectral curve, or  a Riemann surface
on which all Q-functions are defined.
For that we need to know in particular
the analytic  continuations  of \({\bP}_1\) and \( \bP_2\)
through the cut which we denote as \(\tQ_1,\tQ_2\).
Quite expectedly,      \(\tQ_1,\tQ_2\) have an infinite ``ladder''
of cuts, with branch points at \(\pm 2g+ i n\)
for any integer \(n\). To describe completely
the Riemann surface, one should know the analytic continuation through
any of those new cuts, and so on.  One of the main
results of this note is that this complicated cut structure has a stunningly simple algebraic description!

Namely, inspecting the properties of Q-functions of   \cite{Gromov:2011cx} we managed to construct \cite{LargePaper} two additional functions
\(\bP_3\) and \(\bP_4\), again with only one single cut,
such that after the analytic continuation
the four functions \(\tQ_a,\, a=1,2,3,4\), can be expressed as  linear combinations of the
initial \(\bP\)'s
\begin{align}\label{tildeP=-muP}
\tQ_a=-\mu_{ab}\chi^{bc} \bP_c\;,
\end{align}
where $\mu_{ab}$ is a \(4\times 4\)
antisymmetric matrix constrained by
\begin{equation}\label{constraint}
\mu_{12}\mu_{34}-\mu_{13}\mu_{24}+\mu_{14}^2=1\;\;,\;\;
\mu_{23}=\mu_{14}\;,
\end{equation}
and \(\chi\) is an antisymmetric constant \(4\times 4\) matrix
with the only nonzero entries  \(\chi^{23}=\chi^{41}=-\chi^{14}=-\chi^{32}=1\).

Furthermore, the analytic continuation around the branch point \(2g\) of \(\mu_{ab}\) itself, i.e. \(\tilde \mu_{ab}\),  has a very peculiar pseudo-periodicity condition
(see \figref{fig:Cuts})
\begin{equation}\label{pper}
\tilde\mu_{ab}(u)=\mu_{ab}(u+i)\;.
\end{equation}
In other words, if we define a function \(\check\mu\) such that it coincides with \(\mu\)
in the strip \(0<\IM u<1\) but has all its cuts going to infinity then
\eqref{pper} simply tells us that   \(\check\mu\) is a truly \(i\)-periodic function: \(\check\mu_{ab}(u+i)=\check\mu_{ab}\;\).

To close the system of equations on  \(\bP,\mu\)  we  have to find a  condition on \(\mu_{ab}\)  similar to \eqref{tildeP=-muP}.
An important part of it  is already dictated by  \eqref{tildeP=-muP}:
since the branch points are quadratic,
we have \(\tilde{\tilde{\bP}}_a=\bP_a\) which leads to
\(
\bP=-\tilde{\mu}\chi\tilde{ \bP}.
\)
This, together with  \eqref{tildeP=-muP}, gives a set of linear equations fixing the discontinuity
 of the matrix \(\mu\) up to a single unknown factor \(e(u)\):
\(\tilde\mu_{ab}-\mu_{ab}=e(u)( \bP_a\tilde \bP_b-\tilde \bP_a \bP_b)\).
We argue below that \(e(u)=1\)  and hence\begin{equation}\label{tildemu=PPtilde}
\tilde\mu_{ab}-\mu_{ab}=\bP_a\tilde\bP_b-\bP_b \tilde\bP_a\;.
\end{equation}

Eqs. \eqref{tildeP=-muP}, \eqref{pper}, \eqref{tildemu=PPtilde}  represent  our main result   --- a complete non-linear  system of  Riemann-Hilbert equations for the AdS/CFT spectral problem.
They allow us to   walk  through the cuts to any sheet (out of infinite number) of the Riemann surface of the functions \(\bP\) and \(\mu\). In this sense, they give the full description of the spectral curve of the problem. Indeed,  by means of  \eqref{tildeP=-muP} and \eqref{pper}  it is easy to walk through the central cut in \figref{fig:Cuts}. The other cuts are present only in \(\mu\). To define the analytic continuation through them, we use a combination of \eqref{tildeP=-muP}, \eqref{pper} and \eqref{tildemu=PPtilde},
\begin{align}
\label{fun2}
\mu_{ab}(u+i)=\mu_{ab}-\bP_a\bP_e\chi^{ec}\mu_{cb}-\bP_b\bP_e\chi^{ec }\mu_{ac}\,,
\end{align}
which allows to recursively express \(\mu_{ab}(u+i n)\) through \(\mu_{ab}(u)\) and shifted \(\bP\)'s -- the quantities with known monodromies.
We refer to this new formulation of the spectral problem, given by eqs.\eqref{tildeP=-muP}-\eqref{tildemu=PPtilde},  as to the \({\bf P}\mu\) system.

Let us argue now that \(\bP\) and \(\mu\) contain the complete information about the initial Y-system. Indeed,
from \eqref{Tsr} we  {restore}  \(\wT_{1,s}\) for \(s>0\). Furthermore, \(\wT_{2,s}=\wT_{1,1}^{[+s]}\wT_{1,1}^{[-s]},\;\wT_{0,s}=1\) and with a help
of one extra relation \(\wT_{3,2}=\wT_{2,3}\mu_{12}\)
(see \cite{Gromov:2011cx}, where \(\CF^-=\mu_{12}\))  we have  just enough of information to recover any \(\wT_{a,s}\)   using solely the Hirota equation \eqref{T-system}, for any left-right symmetric state. It is just a matter of  elementary algebra to write any Y-function explicitly through \(\bP,\mu\). In particular, we find
\begin{equation}\label{Y11Y22}
Y_{11}Y_{22}=1+\frac{\bP_1\tQ_2-\bP_2\tQ_1}{\mu_{12}}=\frac{\mu_{12}(u+i)}{\mu_{12}(u)}\;.
\end{equation}
We note that the first equality holds for any $e(u)$, but imposing  \cite{Gromov:2011cx} $\tilde Y_{11}\tilde Y_{22}=\frac{1}{Y_{11}Y_{22}}$ we fix $e(u)=1$.
\paragraph*{Asymptotics and charges.} The quantity \eqref{Y11Y22} is known to contain the energy/dimension \(\Delta\)   of the state in its large \(u\) asymptotics  \cite{Gromov:2011cx}: \(\log Y_{11}Y_{22}\simeq i\frac{\Delta-L}{u}\). Similarly, the large \(u\) asymptotics of \(\bP\) and \(\mu\) contains the information about  other conserved charges of the state. In fact,  \(\bP_a^+/\bP_a^-\) is the exact quantum analogue of  the \(S^5\)
eigenvalues of the monodromy matrix \cite{Beisert:2005bm} of classical strings moving in \(AdS_5\times S^5\) and thus \(\bP_a^+/\bP_a^-\simeq 1+M_a/(2iu)\), where \(M_a\) are integer   charges of the global \(SO(6)\) symmetry. For instance, in the \({\mathfrak{sl}}_2\) sector, i.e. for spin \(S\) twist \(L\) operators of the type  \(\Tr Z\nabla_+^S Z^{L-1}\) dual to the string which is point-like in \(S^5\) and moves there with the angular momentum $L$, one has the following asymptotics
\begin{equation}\label{Plarge}
\bP_a\!\simeq\!(A_1u^{-\frac{L}{2}},A_2 u^{-\frac{L+2}{2}},A_3u^{\frac{L}{2}},A_4u^{\frac{L-2}{2}})_a\;.
\end{equation}
Note that at odd \(L\)'s
  \(\bP_a\) have a sign ambiguity   (see   \footnotebis{}).

Next, we also have to specify the asymptotics of \(\mu\).
  Assuming its power-like behavior we immediately get from \eqref{Y11Y22} \(\mu_{12}\simeq u^{\Delta-L}\). To deduce the
asymptotics of the remaining \(\mu\)'s we  consider
\(
\tilde \bP_1=-\mu_{14}\bP_1+\mu_{13}\bP_2-\mu_{12}\bP_3
\)
  and assume that all the terms in the r.h.s. scale in the same way. This gives e.g. \(\mu_{13}\sim\mu_{12}u^{L+1}\sim u^{\Delta+1}\) and, similarly, \((\mu_{14},\mu_{24},\mu_{34})\sim (u^{\Delta},u^{\Delta-1},u^{\Delta+L})\).
This strategy allows one to easily determine the asymptotics for any state even outside of the \(\mathfrak{sl}_2\) sector.

Finally, let us fix the coefficients \(A_i\) in \eqref{Plarge}.  Note that \eqref{fun2}  {becomes at large \(u\)} a homogeneous differential equation
on the \(5\)  {independent} components of \(\mu_{ab}\). By plugging into this equation the asymptotics for \(\mu_{ab}\) and $\bP_a$ we get a \(5\)'th order algebraic equation on \(\Delta\). Its roots are of the form
\((\pm\alpha,\pm\beta,0)\) where \(\alpha,\beta\) are functions of \(A_i\).
 {The root \(\alpha=\Delta\)  reproduces the correct asymptotics of \(\mu_{ab}\),  whereas one can show (see a motivation in discussion) that \(\beta+1=S\) is  the Lorentz spin of the state. By inverting these relations one gets}
\begin{eqnarray}\la{AB}
A_2A_3&=& \frac{[(L-S+2)^2-\Delta^2] [(L+S)^2-\Delta^2)]}{16i L (L+1)}\,,  \nonumber  \\
A_4A_1&=& \frac{[(L+S-2)^2-\Delta^2] [(L-S)^2-\Delta^2]}{16iL (L-1) }\,.
\end{eqnarray}
 Note that \(\Delta\) enters \eqref{AB}  only as \(\Delta^2\), which suggests that the function \(S(\Delta)\)
is  even, as claimed in \cite{Brower:2006ea}. Interestingly,  \(A_i\) enter
only through the products \eqref{AB}, due to a rescaling symmetry
of the \(\bP\mu\) system \cite{LargePaper}.

\paragraph*{Regularity condition.} To single out physical solutions of  the \(\bP\mu\) system we impose the regularity condition: \(\bP\)-s and \(\mu\)-s do not have poles on their defining sheet, and hence,  due to \eqref{tildeP=-muP},\eqref{pper},\eqref{fun2}, on the whole Riemann surface.

\section{Weak coupling}
Let us demonstrate the weak coupling limit for the \(\mathfrak{sl}(2)\) sector.
First,  \eqref{AB} gives an idea about the scaling of
\({\bf P}\)'s at weak coupling: since \(\Delta=L+S+\mathcal{O}(g^2)\),
we see that \(A_2 A_3=\mathcal{O}(g^2)\to 0\) which suggests also that
\({\bf P}_2{\bf P}_3=\mathcal{O}(g^2)\). Hence  at the leading order  \({\bf
  P}_2\simeq 0\) and
 \eqref{fun2} simplifies considerably:  Equations for  \(\mu_{12}\) and \(\mu_{24}\) decouple from the rest.
Excluding \(\mu_{24}\) we get a 2-nd order difference equation  for \(\mu_{12}^+=Q+{\cal O}(g^2)\)
\begin{equation}\label{Baxter}
T\,Q+\frac 1{(\bP_1^-)^2}Q^{[-2]}+\frac 1{(\bP_1^+)^2}Q^{[+2]}=0\,,
\end{equation}
\(T=\frac{\bP_4^+}{\bP_1^+}-\frac{\bP_4^-}{\bP_1^-}-\frac 1{(\bP_1^-)^2}-\frac 1{(\bP_1^+)^2}\),
which is strikingly similar to the Baxter equation
for the Heisenberg spin chain; this analogy goes even further as the zeros
of $\mu_{12}^+$ are indeed {\it exact} Bethe roots \cite{Gromov:2011cx}! To demonstrate the actual equivalence with  Baxter equation one should show that the coefficients in \eq{Baxter} do have the desired analytic properties. Omitting details in this short letter, we only mention that from explicit expression  \cite{Gromov:2011cx} it follows that \({\bf P}_1=A_1u^{-L/2}+\mathcal{O}(g^2)\),
i.e.  the leading order of \({\bf P}_1\) coincides with its large \(u\) asymptotics.
Furthermore, the ratio \({\bf P}_4/{\bf P}_1\) behaves asymptotically as \(u^{L-1}\) and
by construction it has no poles when \(u\neq0\). \(u=0\) is the place where the branch points merge, hence this point is potentially singular.
However, one can advocate that at the leading order \({\bf P}_4/{\bf P}_1\) is also regular at \(u=0\) and hence this ratio is simply a polynomial.
For the same reason of regularity, \(Q\) should be also a polynomial, of degree \(S\) as it follows from the asymptotics of \(\mu_{12}\sim {u^{\Delta-L}}\simeq u^S\).

By standard arguments, zeros of $Q$ should satisfy Bethe equations, which singles out a discrete set of possible $Q's$
and hence of solutions of the $\bP\mu$ system corresponding to the states from the \(\mathfrak{sl}(2)\) sector. For AdS/CFT, we have an additional
zero-momentum constraint $Q(+i/2)/Q(-i/2)=1$ which is due to the cyclicity of trace. The ${\bf P}\mu$ system also encodes this constraint! Indeed,
in the limit $g\to 0$, it is nothing but \eq{pper} evaluated at the branch point $u=2g$ where we used the analyticity condition $\tilde\mu_{ab}(2g)=\mu_{ab}(2g)$.

To compute the one-loop energy we have to compute the large $u$ asymptotics of $\mu_{12}$ to the next order.
From ${\mu_{12}}/Q\sim u^{\Delta-S-L}\simeq 1+(\Delta-S-L)\log u+{\cal O}(g^2))$
we see that we have to find the pre-factor of $\log u$ term.
Such  large $u$ behavior clearly shows that at the next order $\mu_{12}$ can no longer be a polynomial.
Instead, $\mu_{12}$
develops singularities at the collapsing branch cuts $u=i n,\;n\in{\mathbb Z}$  in addition to a modified polynomial part. We denote the singular part of $\mu_{12}^+$ by $R$.
To separate the regular and singular parts we write $\mu_{12}$ in the following way
\beq\label{twelve}
\mu_{12}=\left(\frac{\mu_{12}+\mu_{12}^{++}}{2}\right)+\sqrt{u^2-4g^2}\left[\frac{\mu_{12}-\mu_{12}^{++}}{2\sqrt{u^2-4g^2}}\right],
\eeq
where, due to \eqref{pper}, both expressions inside the brackets have a trivial
monodromy on the cut  $[-2g,2g],$ thus being very smooth near the origin.
The singularity comes solely from the square root factor
whose small $g$ expansion reads: $\sqrt{u^2-4g^2}=u-\frac{2g^2}{u}+\cdots$, which allows us to fix $R^-\simeq g^2 r/u$ with
$r\equiv Q'(\tfrac{i}{2})-Q'(-\tfrac{i}{2})$ in the vicinity of $u=0$. From \eqref{pper} and  \eqref{twelve} we also get $R^+\simeq -g^2 r/u$.

To find all  other possible singularities at \(u\sim i(n+1/2)\) we
notice that in the vicinity of each singularity, \(R\)
must satisfy the same Baxter equation as \(Q\), up to some regular terms. With  the poles of $R(u)$ defined above at \(u=\pm i/2\)
the solution is unique and is given by
\(R(u)=  ig^2 r\frac{Q(u)}{Q(i/2)}\left(\psi(\frac{1}{2}-iu)+\psi(\frac{1}{2}+iu)\right)\).
Now we can  expand \(R\) at large \(u\) to get \(R(u)/Q(u)\simeq \frac{2ir g^2}{Q(i/2)}\log u\) from where we immediately get
\(\Delta=L+S+ \frac{2ir g^2}{Q(i/2)}\), thus reproducing the well known expression for the one-loop dimension
\(
\left.\Delta=L+S+2ig^{2}\partial_u\log\frac{
Q^+}{Q^-}\right|_{u=0}\;.
\)

\section{Cusp Anomalous Dimension}

It was shown in \cite{Drukker:2012de,Correa:2012hh} that the Wilson line with a cusp of an angle \(\phi\)
can be described by essentially the same system of TBA
equations. As a consequence, it can be also studied via the \({\bf P}\mu\) system which turns out to be a very efficient approach, as we are going to demonstrate.
We consider a particular  limit of small \(\phi\). For a more general case, with more details of the derivation, see \cite{grishafedor}.

Whereas  \(\bP\mu\)-equations remain unaltered, it is the large \(u\) behaviour which distinguishes this case from the case of local operators. In particular, one finds  that \(\bP_a\simeq
(A_1  u^{-L+1/2},A_2 u^{-L-1/2},A_3  u^{+L+3/2},A_4 u^{+L+1/2})\) instead of \eqref{Plarge}. Even though \eqref{AB} is not fully applicable now, it appears
to   capture correctly the behaviour of \(\bP_a\) at small \(\phi\): For the case of the vacuum state \(S=0\) and \(\Delta= L+\mathcal{O}(\phi^2)\).
We see that at \(\phi=0\) $A_2A_3\simeq A_4A_1\to 0$ suggesting that to the leading order \(\bP_a=0\).
Hence one gets from \eqref{tildemu=PPtilde} \(\tilde\mu_{ab}=\mu_{ab}\), i.e. \(\mu_{ab}\) has no cuts; it is then just a periodic function as  follows
from \eqref{pper}.

Another specific feature of
this case is that Y-functions have poles
which originate from the boundary dressing phase.
In particular, the product
\eqref{Y11Y22} has simple poles at \(u=in/2\) for any integer \(n\neq 0\) \cite{Correa:2012hh}.
By requiring the regularity of the \(\bP\mu\)-system, we see that in
\eqref{Y11Y22}
the poles can only  originate from  zeros of \(\mu_{12}\).
Hence, \(\mu_{12}\) is a periodic entire function with simple zeros at
\(in/2\). In addition, as Y-functions are even for the vacuum, each
\(\mu_{ab}\) has a certain parity w.r.t. \(u\): for instance
\(\mu_{12}\) is odd and hence it has the form \(\mu_{12}=C\sinh(2\pi u)\).

We have no physical reason to introduce infinite sets of zeros for other \(\mu_{ab}\)'s   and we assume from their periodicity that they are just constants which are further constrained by the parity: \(\mu_{13}=\mu_{24}=0\) because they are odd. Then \(\mu_{34}=0\) and \(\mu_{14}=\pm 1\), in order to satisfy \eqref{constraint}.
A consistent choice of the sign is
\(\mu_{14}=-1\). Then \eqref{tildeP=-muP} gives
\begin{eqnarray}
\label{eq1}&&\tilde{\bf P}_1-{\bf P}_1=-C\sinh(2\pi u){\bf P}_3\;\;,\;\;\tilde{\bf P}_3+{\bf P}_3=0\\
\label{eq2}&&\tilde{\bf P}_2+{\bf P}_2=-C\sinh(2\pi u){\bf P}_4\;\;,\;\;\tilde{\bf P}_4-{\bf P}_4=0\;.
\end{eqnarray}
In what follows we consider for simplicity the case \(L=0\).  The generalization
to arbitrary \(L\) can be done very similarly.
First we notice that in order to cancel the pole in the denominator of
\eqref{Y11Y22} at \(u=0\) we have to assume \(\bP_1\bP_2=0\) at \(u=0\).
If we ``split'' this zero between all \(\bP's\) by introducing a \(\sqrt u\) factor into each of them, we
also ensure a half-integer asymptotics of  \(\bP's\).
From \eqref{eq2} we see that \({\bf P}_4/\sqrt u\) should have no cut and behaves as \(u^0\) at infinity, so it is simply \({\bf P}_4=A_4\sqrt u\).
On the other hand, \({\bf P}_3/\sqrt{u}\) should flip its sign when crossing the cut $[-2g,2g]$ and thus \({\bf P}_3=A_3\sqrt u\sqrt{u^2-4g^2}\).
\({\bf P}_2\) is given from \eqref{eq2} by
the Hilbert transform of \(\sinh(2\pi u)\):
\begin{equation}
\frac{{\bf -P}_2}{C A_4\sqrt u }=\!\!\oint_{-2g}^{2g}\frac{\sqrt{u^2-4g^2}}{\sqrt{v^2-4g^2}}\frac{\sinh(2\pi v)}{4\pi i(v-u)}\!=\!\!\sum_{n=1}^\infty \frac{I_{2n-1}(4\pi g)}{x^{2n-1}},
\end{equation}
where $x(u)$ is defined by $x+\frac 1x=\frac ug$ so we have to set \(A_2=-g C A_4 I_1(4\pi g)\).
Finally, the solution for
 \(\bP_1\) is \(
{\bf P}_1=-\frac{A_3}{A_4}\sqrt{u^2-4g^2}{\bf P}_2+
(A_1+A_3 A_2/A_4)\sqrt u
\).
Now, we introduce  \(\phi\) by requiring that \(1+Y_{11}\simeq -\frac{\phi^2}{2}\) for \(u\to \infty\)
and find the energy from \(Y_{11}Y_{22}-1\simeq 2i\Delta/u\) (note an extra two in this equation which is due to the open boundary conditions). We notice that to match these
expansions we
should first assume \(A_1A_4=A_2A_3\) as otherwise $Y_{11}Y_{22}-1$ would grow
linearly. Then the first condition gives
\(-\frac{\phi^2}{2}=\frac{i}{2}A_1A_4\) and from the second
\begin{align}
{\Delta}=&
-\phi^2 g^2 \left(1-\frac{I_3(4\pi g)}{I_1(4\pi g)}\right)\,
\end{align}
 -- the same result as found from localization in \cite{Correa:2012at,Fiol:2012sg} or using TBA/FiNLIE approach in \cite{Gromov:2012eu}.

\section{Discussion}
In this letter we formulated the \(\bP\mu\) system -- a new way to
describe the AdS/CFT spectrum. This system seems to be well
suited
to address  various long-standing open
problems, including a systematic study of strong coupling of short operators and the BFKL regime.
We can also benefit from it for a systematic weak coupling expansion and
the study of  Wilson loops.

At the same time, the \(\bP\mu\) system provides a new conceptual insight into
the AdS/CFT integrability.
In particular, the present \(\bP\mu\)-system, with
\(\bP_a^+/\bP_a^-\) corresponding  to the \(S^5\) eigenvalues
of the quasiclassical  monodromy matrix, is  the perfect counterpart of  the \(\bQ\omega\)-system to be described in \cite{LargePaper}:
the four fundamental fermionic Q-functions \(\bQ_{\hat a}\) have only one long
cut \((-\infty,-2g]\cup[2g,\infty)\) and their
monodromies are expressed through a  \(4\times 4\)
matrix \(\omega\) (periodic on the sheet with short cuts). We believe that \(\bQ_{\hat a}^+/\bQ_{\hat a}^-\)
correspond to the \(AdS_5\) eigenvalues of \(\Omega\).

These two systems are related by linear relations of the type \(\mu_{ab}=\bQ_{ab\hat
a\hat b}\omega^{\hat a\hat b}\), which allowed us  to  explicit the  Lorenz spin \(S\) dependence of the coefficients in \eqref{AB} and thus to close the \(\bP\mu\)-system on
itself \cite{LargePaper}. In addition, the symmetry between these two systems would
a priori allow to interchange the role of \(\bP_a\) and \(\bQ_{\hat
a}\). One interesting application of
this is the possibility
to construct the ``physical T-hook'' -- where the Y- and T-systems have the same
algebraic formulation as in the original mirror T-hook, but all cuts are
short instead. At weak coupling, short cuts collapse and we expect the
1-loop physical T-functions to be the eigenvalues of  transfer matrices of
the  \(\mathfrak{psu}(2,2|4)\) XXX spin chain \cite{Beisert:2005di}. We describe this construction and  the full derivation
of the \(\bP\mu\) system in our future work \cite{LargePaper}.
 The exact physical T-functions seem to  represent the eigenvalues of, yet to be constructed, T-operators of  all-loop N=4 SYM spin chain.

Finally, let us note that the  monodromy around a branch point corresponds to the
crossing transformation and one can speculate that \eqref{tildeP=-muP}
is a crossing QQ-relation,  related to the bubble Y-system
of \cite{Alday:2009dv,Alday:2010vh}.
  \begin{acknowledgments}
\section*{Acknowledgments}
\label{sec:acknowledgments}
We thank M. Alfimov, G. Sizov and F. Levkovich-Maslyuk for discussions. The work of V.K. is  supported by the ANR grant StrongInt (BLANC-
SIMI- 4-2011), by RFBR grant 11-02-01220 and by the ESF grant HOLOGRAV-09-RNP- 092. V.K. also appreciates
the support of Institut Universitaire de France. The work of S.L. is supported
by the ERC Advanced grant No.290456.
The research of N.G. and V.K. leading to these results has received funding from the
People Programme (Marie Curie Actions) of the European Union's Seventh
Framework Programme FP7/2007-2013/ under REA Grant Agreement No 317089.
\end{acknowledgments}

\ifbibtexelse{
\bibliography{biblio.bib}
}
{

}

\end{document}